\documentclass[preprint,12pt,3p]{elsarticle}
\usepackage{latexsym}
\usepackage{dcolumn}
\usepackage{supertabular,multirow}
\usepackage{epsfig}
\usepackage{bm}
\usepackage{subfigure}
\usepackage{color}
\usepackage{xurl}
\usepackage{tikz}
\usepackage{amssymb,amsfonts,amsmath}
\usepackage{euscript}
\usepackage{graphics}
\usepackage{setspace}
\usepackage{graphicx}
\usepackage{hyperref}
\usepackage{xcolor}
\usepackage{siunitx}
\usepackage[normalem]{ulem}
\usepackage[colorinlistoftodos]{todonotes}
\usepackage{textgreek}

\graphicspath{{./}, {Figs/}}

\definecolor{darkgreen}{rgb}{0, 0.5, 0.05}
\definecolor{orange}{rgb}{1, 0.35, 0.05}

\newcommand{\deleted}[1]{}

\newcommand{\eqn}[1]{\begin{align}#1\end{align}}
\newcommand{\bs}[1]{\boldsymbol{#1}}

\newcommand{\pare}[1]{\left( #1 \right) }
\newcommand{\corchete}[1]{\left[ #1 \right]}
\newcommand{\fr}[2]{\frac{#1}{#2}}
\newcommand{\wtil}[1]{\widetilde{#1}}
\newcommand{\mc}[1]{\mathcal{#1}}
\newcommand{\avg}[1]{\langle #1 \rangle}

\def\dd{\mathrm{d}}  

\def\bna{\bs{\nabla}}

\def\bB{\bs{B}}
\def\bD{\bs{D}}
\def\be{\bs{e}}
\def\bbf{\bs{f}}
\def\bF{\bs{F}}

\def\bI{\bs{I}}

\def\bM{\bs{M}}

\def\bn{\bs{n}}

\def\bP{\bs{P}}
\def\bK{\bs{K}}
\def\bq{\bs{q}}

\def\br{\bs{r}}

\def\bu{\bs{u}}
\def\bU{\bs{U}}
\def\bv{\bs{v}}

\def\bx{\bs{x}}

\def\by{\bs{y}}

\def\bzero{\bs{0}}

\def\btau{\bs{\tau}}
\def\bxi{\bs{\xi}}
\def\bomega{\bs{\omega}}

\def\blambda{\bs{\lambda}}

\def\bmM{\bs{\mc{M}}}

\def\bmD{\bs{\mc{D}}}

\def\mcB{\mc{B}}

\def\bmK{\bs{\mc{K}}}

\def\bmM{\bs{\mc{M}}}

\def\bmP{\bs{\mc{P}}}

\def\bmI{\bs{\mc{I}}}

\begin{document}

\title{Simulating non-Brownian suspensions with non-homogeneous Navier slip boundary conditions}

\author[mymainaddress1,mymainaddress2]{Daniela Moreno-Chaparro}
\author[mymainaddress1]{Florencio Balboa Usabiaga}
\author[mymainaddress1]{Nicolas Moreno}
\author[mymainaddress1,mymainaddress2,mymainaddress4]{Marco Ellero}

\address[mymainaddress1]{BCAM - Basque Center for Applied Mathematics, Mazarredo 14, Bilbao, E48009, Basque Country - Spain}
\address[mymainaddress2]{University of the Basque Country - Euskal Herriko Unibertsitatea, Leioa, E48940, Basque Country - Spain}
\address[mymainaddress4]{Complex Fluids Research Group, Department of Chemical Engineering, Faculty of Science and Engineering, Swansea University, Swansea SA1 8EN, United Kingdom}

\date{\today}

\begin{abstract}

  Fluid-structure interactions are commonly modeled using no-slip boundary conditions.
  However, small deviations from these conditions can significantly alter the dynamics of suspensions and particles, especially at the micro and nano scales.
  This work presents a robust implicit solvent method for simulating non-colloidal suspensions with non-homogeneous Navier slip boundary conditions.
  Our approach is based on a regularized boundary integral formulation, enabling accurate and efficient computation of hydrodynamic interactions.
  This makes the method well-suited for large-scale simulations.
  We validate the method by comparing computed drag forces on homogeneous and Janus particles with analytical results.
  Additionally, we consider the effective viscosity of suspensions with varying slip lengths, benchmarking against available analytical no-slip and partial-slip theories.
  
\end{abstract}
\maketitle

\section{Introduction}
\label{sec:Intro}

The study of suspensions dynamics often assume the no-slip boundary conditions (BCs) on the particles surfaces for practical purposes.
The no-slip condition states that the fluid velocity at a solid boundary matches the velocity of the boundary itself.
This approximation has been successfully used to explain many fluid dynamics phenomena, such as the dynamics of particles in flow and the rheology of suspensions \cite{Jeffery1922, Happel1983}.
However, the no-slip BCs are not always satisfied, and even small deviations can significantly influence the hydrodynamics of small particles \cite{Bocquet2007,Neto2005}.
The no-slip BCs can be indeed generalized by the partial slip, or Navier BCs models, which introduce the concept of slip length \cite{Neto2005}, being the distance across the boundary at which the linearly extrapolated fluid velocity would reach zero. For macroscopic objects, even when the slip length is nonzero, it is typically much smaller than the objects' characteristic dimensions. As a result, the no-slip BCs provide a very good approximation in most cases. However, for micro- and nanoparticles, even small slip lengths can significantly influence particle dynamics. Consequently, partial slip BCs become important when dealing with systems approaching the molecular scale \cite{Shuvo2024}. Partial slip BCs can influence the suspension's viscosity \cite{Taylor1932,Oldroyd1953,Allison1999,Kamal2023} and overall rheological behavior, especially in systems where interparticle interactions dominate \cite{Vazquez2018}. Moreover, it has been suggested that the slip length may vary with the local shear rate, indicating a potential coupling between partial slip BCs and flow conditions \cite{Vazquez2018,Kroupa2017,Housiadas2020}.

Several experimental techniques have been developed to quantify slip lengths across various systems. One method involves tracking rough particles, where surface roughness mimics the effects of slip length \cite{Scherrer2024, Hsu2018, Niggel2023}. Common experimental approaches for measuring slip length include surface force apparatus (SFA), atomic force microscopy (AFM), micro-particle image velocimetry (\textmu-PIV) \cite{Shuvo2024}, suspended microchannel resonator (SMR), dynamic quartz crystal microbalance (QCM-D), and hybrid graphene/silica nanochannel techniques \cite{Shuvo2024}. Other indirect techniques such as pressure drop versus flow rate, drainage force measurements, and streaming potential analysis have also been employed \cite{Lauga2007}.
Moreover, non-uniform slip conditions-such as those found in Janus particles with asymmetric surface properties—have been studied both experimentally \cite{Lattuada2011} and theoretically \cite{Swan2008, Ramachandran2009, Premlata2019, Premlata2022, Biswas2023}.

In addition to experimental and theoretical studies, numerical methods have been used to investigate systems with partial slip BCs.
These methods can be broadly categorized into explicit and implicit solvent methods.
Explicit solvent methods directly resolve the fluid flow and include molecular dynamics (MD) \cite{Lauga2007, Kamal2020, Pelaez2025},
smoothed particle hydrodynamics (SPH) \cite{colagrossi2003}, smoothed dissipative particle dynamics (SDPD) \cite{Cai2023},
and dissipative particle dynamics (DPD) \cite{Xu2019}, immersed boundary method (IBM) \cite{Peskin2002}, among others.
The explicit discretization of the fluid makes these methods versatile, as they can deal with, for example,
non-constant viscosity \cite{Krishnan2017, VazquezQuesada2019} but also computationally demanding.
In contrast, implicit solvent methods do not solve for the flow field directly.
Instead, they use the Green's functions of the Stokes equations to compute the velocities of solute particles.
These methods are limited to the study of suspensions in the Stokes regime.
However, with this methods only the particles surfaces are discretized  which reduces memory requirements considerably.
Some implicit analytical methodologies have been also developed to include partial slip BCs \cite{Luo2008, Sun2013, Camargo2019, Kamal2023, Biswas2023}.

This work presents a robust implicit solvent method for modeling particles with partial slip BCs, capable of simulating complex and long-range interactions in particle suspensions. 
Additionally, the particles can feature non-homogeneous slip lengths along their surfaces. Our method employs a regularized boundary integral formulation to implement partial slip BCs and compute hydrodynamic tractions and particle velocities \cite{Smith2021}. 
The approach achieves high accuracy, with only a few percent relative errors, even with a modest discretization per particle.
Thus, it is appropriate for simulating large-scale particulate systems.

We validated our method by modeling spherical particles with uniform partial slip and no-slip conditions.
We then study Janus particles, which exhibit asymmetric slip distributions over their surfaces, leading to complex anisotropic hydrodynamic behavior.
Physically, a spherical particle with a uniform slip length tends to move along the direction of an applied force.
However, in the case of non-uniform slip distributions, such as with Janus particles,
the particle may move with an angle respect to the apply force and experience also a torque\cite{Keh1985}.
This effect naturally arises since the increase in the slip length reduces locally the fluid's resistance,
producing unbalance and asymmetric forces that influence the particle's translational and rotational motion, even for spherical geometries \cite{Swan2008}.

The structure of the paper is the following: We begin in Section \ref{sec:formulation} by introducing the continuum formulation for modeling particles with partial slip BCs \cite{Kamal2021},
including the definition of single-layer and double-layer operators. Section \ref{sec:discretization} details the discretization of the governing equations and the use of a preconditioned iterative solver to efficiently solve the Stokes problem. In Section \ref{sec:validation}, we validate the method by computing hydrodynamic drag on particles with both homogeneous and heterogeneous slip distributions over a wide range of slip lengths, comparing our numerical results with analytical predictions \cite{Taylor1932,Happel1983,Sierou2001,Housiadas2020}. Finally, we explore how slip length affects the relative viscosity of suspensions at different volume fractions, observing a decrease in viscosity with increasing slip length for a given volume fraction.

\section{Continuum formulation}
\label{sec:formulation}

In this section we describe the formulation to model suspended rigid particles with partial slip BCs.
We consider a collection of $M$ arbitrarily shaped particles, $\{\mcB_m\}_{m=1}^M$,
immersed in a viscous dominated flow.
At negligible Reynolds number the flow is governed by the Stokes equations
\eqn{
  \label{eq:Stokes}
  -\bna p + \eta \bna^2 \bv &= \bzero, \\
  \bna \cdot \bv &= 0,
}
where $\bv$ and $p$ are the flow velocity and pressure and $\eta$ the constant fluid viscosity.
The BCs on the particles surfaces are model as a \emph{slip} condition
\eqn{
  \label{eq:slip_condition}
  \bv(\br) = \bu_m + \bomega_m \times (\br - \bq_m) + \bu_s(\br)  \text{ for $\br$ on } \partial \mcB_m,
}
where $\bu_m$ and $\bomega_m$ are the linear and angular velocities of the particle $m$ around its tracking point (e.g.\ its center) $\bq_m$
and $\bu_s$ is the slip velocity.
For $\bu_s=\bzero$ the standard no-slip conditions are recovered.
Instead, we use $\bu_s$ to impose mixed Dirichlet-Neumann BCs on the flow velocity.
We model the particle as impermeable particles, thus, the flow obeys the no-penetration boundary condition or,
equivalently, the normal component of the slip vanishes, i.e.\ $\bu_s(\br) \cdot \bn(\br) = 0$ with $\bn(\br)$ being the unit normal to the particle surface at $\br$.
The tangential component of the slip is modeled with Navier BCs, i.e.\ the
magnitude of the slip is proportional to the local hydrodynamic traction 
\eqn{
  \label{eq:xi_us}
  \xi (\bI - \bn \bn^T) \bu_s = -(\bI - \bn \bn^T) \blambda,}
where $\blambda$ is the traction exerted on the fluid and $\xi=\xi(\br)$ the sliding friction coefficient that can vary over the particle surface
\cite{Happel1983, Bocquet2007, Swan2008}.
We can write $\xi=\eta / \ell$, where $\eta$ is the fluid viscosity and $\ell=\ell(\br)$ a characteristic length, the so-called \emph{slip length}.
In the limit $\xi\rightarrow\infty$ (equivalently $\ell=0$) the no-slip condition is recovered;
the free-slip condition is achieved for $\xi=0$ ($\ell \rightarrow \infty$) \cite{Bocquet2007}.
Both conditions on the slip velocity can be combined into one equation
\eqn{
  \label{eq:us}
  \xi \bu_s = -\bmP \blambda = -\pare{\bI - \bn \bn^T} \blambda,
}
where $\bmP = (\bI - \bn \bn^T)$ is the projector operator into the particle surface tangential space.

The equations are closed by the balance of force and torque, that is, the hydrodynamic 
traction on the particles balances the external forces, $\bbf_m$, and torques, $\btau_m$, applied to them 
\eqn{
  \label{eq:balanceF_continuum}
  \int_{\partial \mcB_m} \blambda(\br) \,\dd S_{\br} = \bbf_m, \\
  \label{eq:balanceT_continuum}
  \int_{\partial \mcB_m} (\br - \bq_m) \times \blambda(\br) \,\dd S_{\br} = \btau_m.
}
We make use of an integral formulation to solve the Stokes equations \cite{Pozrikidis1992}.
First, we introduce the single layer operator, $\bmM$, acting on the hydrodynamic traction $\blambda$,
and the Stokes double layer operator, $\bmD$, acting on the fluid velocity
\eqn{
  \label{eq:operators_Stokes} 
  \pare{{\bmM} \blambda}_i(\bx) &= \int_{\partial \mcB} \mathbb{M}_{ij}(\br', \br'')  \lambda_j(\by) \; \dd S_y = \int_{\partial \mcB} \fr{1}{8\pi \eta r} \pare{\delta_{ij} + \fr{r_i r_j}{r^2}} \lambda_j(\by) \; \dd S_y, \\  
  \pare{\bmD \bv}_i(\bx) &= \int_{\partial \mcB}  \mathbb{D}_{ijk}(\br', \br'') n_k(\by) v_j(\by) \; \dd S_y, = \int_{\partial \mcB} -\fr{3}{4\pi} \fr{r_i r_j r_k}{r^5} n_k(\by) v_j(\by) \; \dd S_y,
}
in an unbounded domain.
Using these operators the double layer formulation for the fluid velocity in the particle surfaces is \cite{Pozrikidis1992}
\eqn{
  \label{eq:double_layer}
  \pare{\fr{1}{2}\bmI + \bmD} \bv(\br) = (\bmM \blambda)(\br) \;\;\; \text{for }\br \text{ on } \partial \mcB_m,
}
where $\bmI$ is the identity operator.
Replacing the slip condition, Eq.\ \ref{eq:slip_condition}, one obtains \cite{Kamal2020, Kamal2021}
\eqn{
  \label{eq:slip_integral}
  \pare{\fr{1}{2}\bmI + \bmD }\pare{\bu_m + \bomega_m \times (\br - \bq_m) + \bu_s(\br)} = (\bmM \blambda)(\br) \;\;\; \text{for }\br \text{ on } \partial \mcB_m.
}
The equations \eqref{eq:us}-\eqref{eq:balanceT_continuum} and \eqref{eq:slip_integral} form a linear system for the unknown
particle velocities, $\bu_p$ and $\bomega_p$, hydrodynamic traction, $\blambda$, and surface slip $\bu_s$.
In the next section we introduce a specific discretization and a method to solve the corresponding linear system.

\section{Discretization}
\label{sec:discretization}
\begin{figure}
  \centering
  \includegraphics[width=0.95 \textwidth]{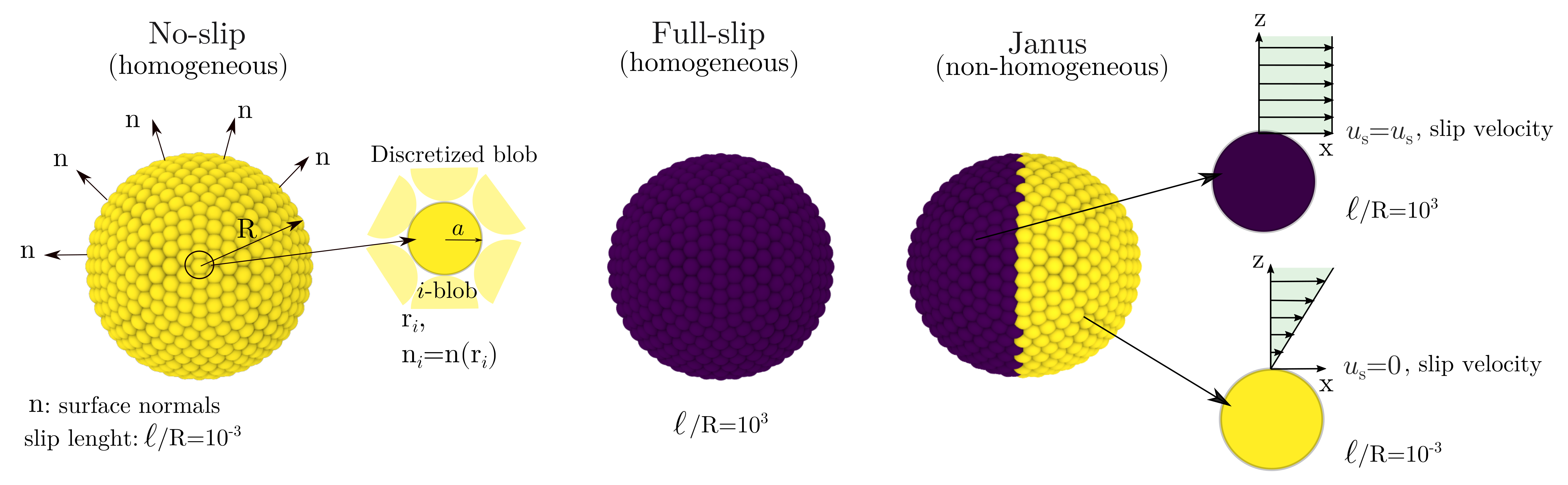}
  \caption{Examples of discretized spherical particles with different slip distributions: homogeneous no-slip, homogeneous full-slip and Janus slip distribution
    which we model with slip lengths $\ell / R = 10^{-3}$ and $\ell / R = 10^{3}$ respectively, where $R$ is the particle radius.
    The surface normals are denoted as $\text{n}$.
  }
  \label{fig:models}
\end{figure}

We discretize the surface of the particles with markers or \emph{blobs} of radius $a$ with position $\br_i$.
At each blob the surface normal, $\bn_i=\bn(\br_i)$, is also defined, see Fig.\ \ref{fig:models}.
After the discretization, the balance of force and torque, Eqs.\ \eqref{eq:balanceF_continuum}-\eqref{eq:balanceT_continuum},
become sums over the blobs of each body $m$ 
\eqn{
  \label{eq:balance_f_blobs}
  \sum_{j \in \mcB_m} \blambda_j &= \bbf_m, \\
  \label{eq:balance_tau_blobs}
  \sum_{j \in \mcB_m} (\br_j - \bq_m) \times \blambda_j &= \btau_m.
}
We have included the quadrature weights into the traction, $\blambda_i$,
so they represent finite forces instead of density forces as in Sec.\ \ref{sec:formulation}.
The BCs for the partial slip, Eq.\ \eqref{eq:us}, are evaluated at each blob, that is
\eqn{
  \label{eq:us_discrete}
  \xi_i \bu_{s,i} = -\bP_{i} \blambda_i = -w_i^{-1} \pare{\bI - \bn_i \bn_i^T} \blambda_i,
}
where $w_i$, the quadrature weight of blob $i$, accounts for the fact that the Navier boundary condition is written in terms of density forces.

The partial slip condition, Eq.\ \eqref{eq:slip_integral}, is also evaluated on each blob $i$
\eqn{
  \label{eq:no-slip_blobs}
  \sum_{j=1}^{N_b} \bM_{ij} \blambda_j &= \sum_{j=1}^{N_b} \pare{\frac{1}{2}\bI_{ij}+\bD_{ij}} \pare{\bu_m + \bomega_m \times (\br_j - \bq_m) + \bu_{s,j}}, 
}
where $\bM$ and $\bD$ are discretizations of the single layer and double layer Stokes kernels and
$\bI_{ij}$ is a $3\times 3$ identity matrix for $i=j$ and zero otherwise.

The method uses a regularized version of the Stokes Green's function, the Rotne-Prager mobility, which can be defined for non-overlapping blobs as \cite{Rotne1969,Wajnryb2013}
\eqn{
  \label{eq:RPY}
  \bM_{ij} = \bM(\br_i, \br_j)  &=  \left. \corchete{\bI + \fr{a^2}{6}\bna^2_{\br'}} \corchete{\bI + \fr{a^2}{6}\bna^2_{\br''}} \mathbb{M}(\br', \br'')\right|_{\br'=\br_i}^{\br''=\br_j}.
}
For overlapping blobs it is possible to remove the singularity associated with the divergence of the Stokes kernels, see Ref.\ \cite{Wajnryb2013}.
For reliable hydrodynamics interactions, is recommend a minimum distance of one blob radius $r_o=a$, such that the blobs do not overlap. 
For the double layer we use the same regularization procedure shown in Eq.\ \eqref{eq:RPY}.

The four equations \eqref{eq:balance_f_blobs}-\eqref{eq:no-slip_blobs} form a linear system for the
particles velocities, tractions and slip velocities.
This linear system can be written in compact notation with the help of the geometric matrix $\bK$,
which transform rigid body velocities into surface velocities
\eqn{
  \label{eq:K}
  \bK\begin{bmatrix}
  \bu_m \\
  \bomega_m 
  \end{bmatrix} = \bu_m + \bomega_m \times (\br_j - \bq_m) \text{ for } j \in \mcB_m.
}
Then, combining the four equations \eqref{eq:balance_f_blobs}-\eqref{eq:no-slip_blobs} and using \eqref{eq:K} gives the linear system
\begin{equation}
  \label{eq:linear_system}
    \begin{bmatrix}
      \bM & - \fr{1}{2}\bK - \bD\bK & -\fr{1}{2} \bI-\bD \\
    -\bK^T & \bzero & \bzero \\
    \bxi^{-1}\bP &  \bzero & \bI \\
    \end{bmatrix}
    \begin{bmatrix}
      \blambda\\
      \bU \\
      \bu_s \\
    \end{bmatrix} =
    \begin{bmatrix}
      \bzero \\
    -\bF \\
    \bzero \\
    \end{bmatrix},
\end{equation}
where the velocities and force-torque on the particles are collected on the vectors $\bU=\{\bu_m, \bomega_m\}_{m=1}^M$
and $\bF=\{\bbf_m, \btau_m\}_{m=1}^M$ and the unscripted vectors $\blambda$ and $\bu_s$ contain the traction and slip.
The unscripted matrices are defined accordingly.

\subsection{Numerical solver}
\label{sec:solver}

\begin{figure}
  \centering
  \includegraphics[width=1 \textwidth]{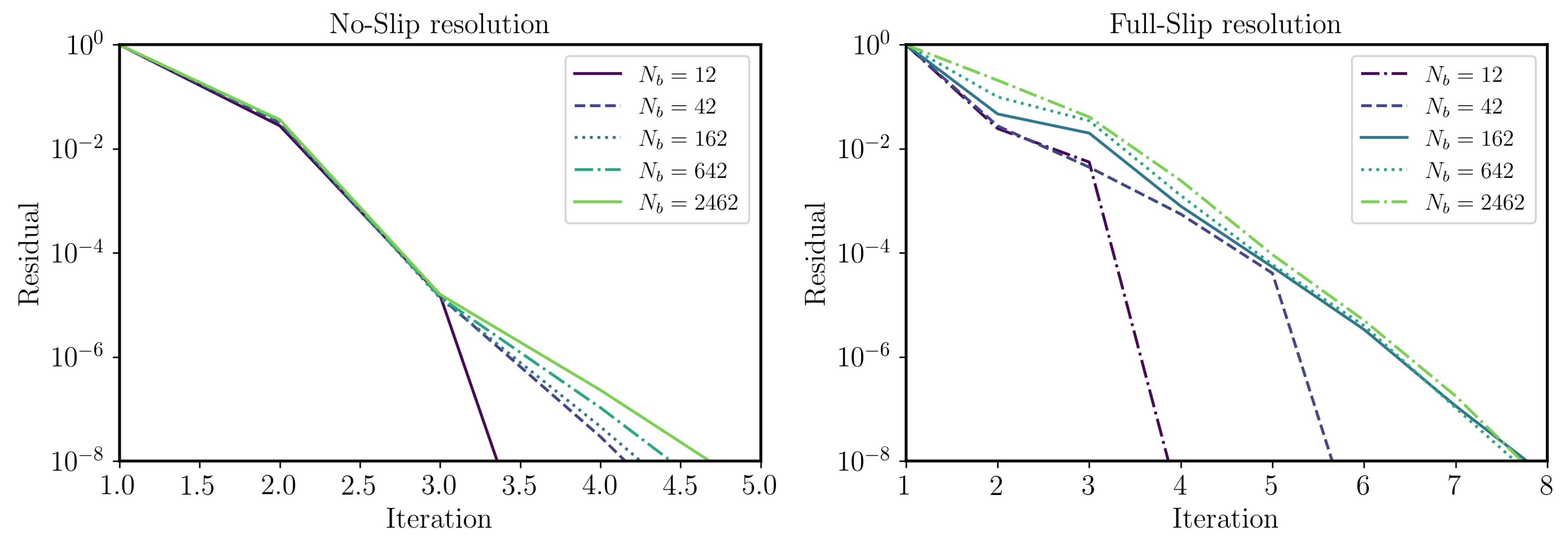}
  \caption{Convergence of preconditioned GMRES to solve the mobility problem Eq.\ \eqref{eq:linear_system}
    for a single particle discretized with a different number of blobs.
    The figure shows results for no-slip (left) and full-slip (right) BCs.}
  \label{fig:convergence_one_shell}
\end{figure}

\begin{figure}
  \centering
  \includegraphics[width=1 \textwidth]{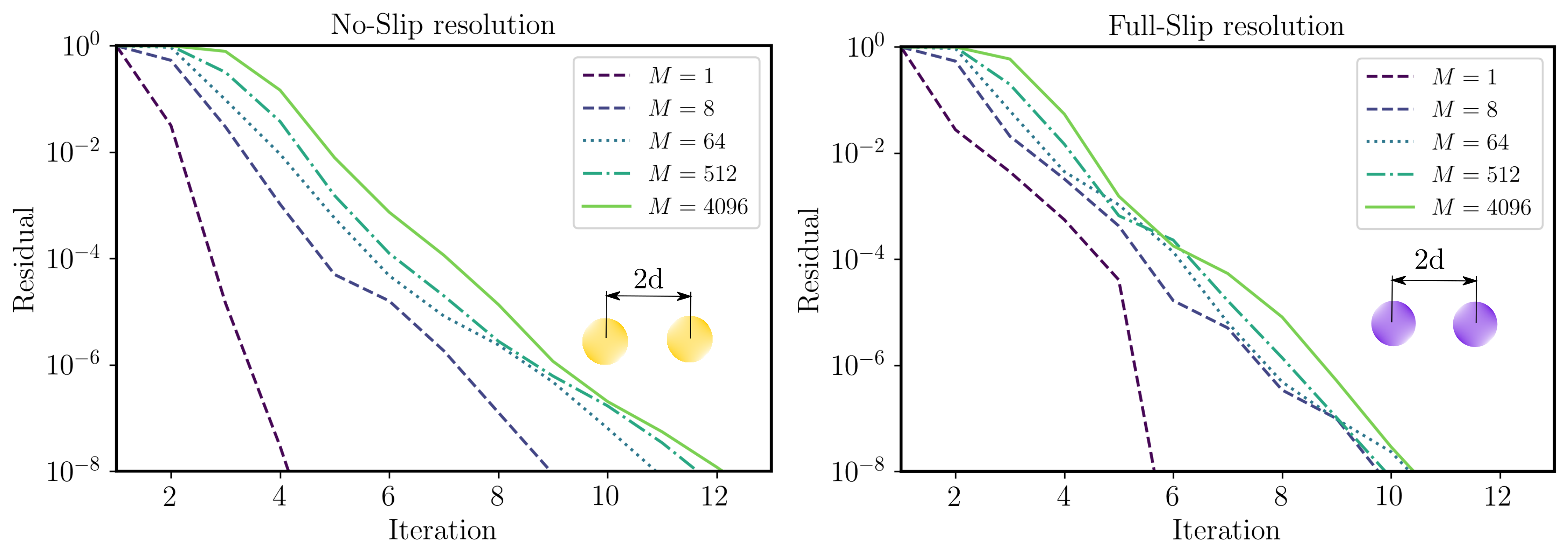}
  \caption{GMRES convergence to solve the mobility problem Eq.\ \eqref{eq:linear_system} for $\bM$ particles, from 1 to 4096,
    forming a simple cubic lattice with a distance between first neighbors equal to two particle diameters.
    The particles are discretized with $N_b$ = 42 blobs. 
  }
  \label{fig:convergence_many_shells}
\end{figure}

The linear system Eq.\ \eqref{eq:linear_system} can be large for many-particle suspensions, therefore, using a direct solver is computationally unfeasible.
Instead, we solve it with an iterative method.
Since Eq.\ \eqref{eq:linear_system} is non-symmetric we employ a preconditioned GMRES solver which has been shown to be a very efficient for similar Stokes problems
\cite{Usabiaga2016, Delmotte2024}.

To converge within a moderate number of iterations it is necessary to employ a preconditioner which approximates the inverse of Eq.\ \eqref{eq:linear_system}.
Since the preconditioner speeds up the convergence but it does not modify the solution of the linear system we can make a number of approximation
so it can be applied efficiently.
First, we note that in the continuum limit the action of the double layer on rigid body velocities is equivalent to applying half the identity operator \cite{Pozrikidis1992}.
Thus, on the preconditioner we approximate $\pare{\fr{1}{2}\bI + \bD}\bK \approx \bK$.
Second, in the mobility matrix we neglect the interaction between blobs belonging to different bodies,
i.e.\ we  replace the mobility matrix $\bM$ by its block-diagonal approximation
$\wtil{\bM}^{(pq)} = \delta_{pq} \bM^{(pp)}$ where the superscripts $p$ and $q$ indicate the block of blobs belonging to body $p$ or $q$ respectively.
Finally, we simply neglect the action of the double layer on the slip velocity.
Applying the preconditioner derived with these approximations is equivalent to solve the system
\begin{equation}
  \begin{bmatrix} 
    \wtil{\bM} & -\bK & -\bI \\
    -\bK^T & \bzero & \bzero \\
    \xi^{-1}\bP & \bzero & \bI \\
  \end{bmatrix}
  \begin{bmatrix}
    \blambda\\
    \bU \\
    \bu_s \\
  \end{bmatrix} =
  \begin{bmatrix}
    \bu_0\\
    -\bF \\
    \bB \\
  \end{bmatrix}.
\end{equation}
Note that it is necessary to include $\bu_0$ and $\bB$ in the right hand side because during the GMRES iterations those terms are not guarantee to be zero
even if they are zero in the original Eq.\ \eqref{eq:linear_system}.
Since the preconditioner does not include many body interactions it can be solved easily for each body.
The solution of the preconditioner is
\eqn{
  \bU &= \corchete{\bK^T \pare{\wtil{\bM} + \xi^{-1} \bP}^{-1}\bK}^{-1}\pare{\bF - \bK^T \pare{\wtil{\bM} + \xi^{-1} \bP}^{-1}(\bu_0 + \bB)}, \\
  \blambda &= \pare{\wtil{\bM} + \xi^{-1} \bP}^{-1} \corchete{\bK \bU + \bu_0 + \bB}, \\
  \bu_s &= -\bxi^{-1} \bP \blambda + \bB.
}
We can see that the only expensive calculation to compute the action of the preconditioner correspond to
computing the inverse $\pare{\wtil{\bM}^{(pp)} + \xi^{-1} \bP^{(pp)}}^{-1}$ for each body $p$.
Since these matrices are defined for a single body they are not too big and their inverses can be computed with dense algebra methods,
in particular, we use a Cholesky factorization.

We show in Fig.\ \ref{fig:convergence_one_shell} the convergence of GMRES to solve the mobility problem Eq.\ \eqref{eq:linear_system}
for a single particle discretized with different number of blobs.
We see that regardless of the discretization and the slip length the solver converges quickly, within a small number of iterations.
In Fig.\ \ref{fig:convergence_many_shells} we show the convergence for particles, discretized with $42$ blobs,
forming a simple cubic lattice with distances between first neighbors of two particle diameters.
The figure shows that the convergence depends very weakly on the number of particles even for very large systems.

\section{Numerical validation}
\label{sec:validation}
In this section, we validate the scheme by computing the drag on spherical particles with different slip lengths and slip distributions.
Each sphere, of \emph{geometric} radius $R_g=1$, is discretized with $N_b$ blobs of radius $a$.
We set the blob radius to $a=d/2$ where $d$ is the smallest distances between blobs so nearest neighbors touch each other.
We will see that for a finite number of blobs the particle responses to the flow like
a sphere with an effective hydrodynamic radius $R_h > R_g$.
However $R_h$ converges to $R_g$ as the number of blobs increases.
Figure \ref{fig:models} illustrates some of the discretized particles used in this section.

\subsection{Mobility}
\label{sec:mobility}

First, we measure the effective hydrodynamic radius of a spherical particle with different resolutions to determine the scheme accuracy.
Here we consider two cases,  no-slip and full-slip BCs, which we implement using the slip lengths $\ell/R_g=10^{-6}$ and $\ell/R_g=10^{3}$ respectively.
We apply a force or torque to the particles and define the effective radii as the translational hydrodynamic radius $R_h = \frac{f_p}{\alpha\pi\eta u_p}$,
and the rotational radius as $R_\tau = \pare{\frac{\tau_p}{8\pi\eta\omega_p}}^{1/3}$,
where $\alpha=6$ or $\alpha=4$ for the no-slip and full-slip cases respectively.
The measured radii are shown for several discretizations in Table \ref{tab:my-table2}.
We note that for the full-slip BCs the rotational hydrodynamic radius is not well defined because no drag acts on a rotating particle.
Therefore, in the full-slip case we only report the translational hydrodynamic radius.
In all cases the values converge to $R_g$  with higher resolutions.


\begin{table}[h]
\centering
\begin{tabular}{@{}|c|c|c|c|c|@{}}
\hline
\begin{tabular}[c]{@{}c@{}}Number of blobs \\ $N_b$\end{tabular} &
\begin{tabular}[c]{@{}c@{}}Blob radius \\ $a/R_g$\end{tabular} &
\begin{tabular}[c]{@{}c@{}}No-Slip \\ $R_h/R_g$\end{tabular} &
\begin{tabular}[c]{@{}c@{}}No-Slip\\ $R_\tau/R_g$\end{tabular} &
\begin{tabular}[c]{@{}c@{}}Full-Slip \\ $R_h/R_g$\end{tabular} \\ \hline
12   & 0.5257 & 1.0819 & 1.0826 & 1.0345 \\ \hline
42   & 0.2733 & 1.0321 & 1.0338 & 1.0046 \\ \hline
162  & 0.1380 & 1.0086 & 1.0152 & 0.9942 \\ \hline
642  & 0.06914 & 1.0020 & 1.0072 & 0.9944 \\ \hline
2562 & 0.03459 & 1.0003 & 1.0035 & 0.9964 \\ \hline
\end{tabular}
\caption{Effective translational and rotational radii for spheres discretized with different number of blobs.
  Results for sphere with homogeneous no-slip and full-slip BCs.
}
\label{tab:my-table2}
\end{table}

Table \ref{tab:my-table2} shows that both the translational and rotational hydrodynamic radius are close for all discretizations.
In practical applications, if we wanted to study particles of radius $R$ we would set the hydrodynamic radius
to the target radius of the particles under study $R_h = R$ (thus, $R_g < R$ for finite resolutions).
This allows us to get results accurate up to a few percent using coarse discretizations.
For complex shaped particles one can compute the particle mobility to high accuracy and then optimize a low resolution grid to match the
high accuracy computed mobility \cite{Broms2023, Delmotte2024}

\subsection{Drag}
\label{sec:drag}

\begin{figure}
\centering
\includegraphics[width=0.49 \textwidth]{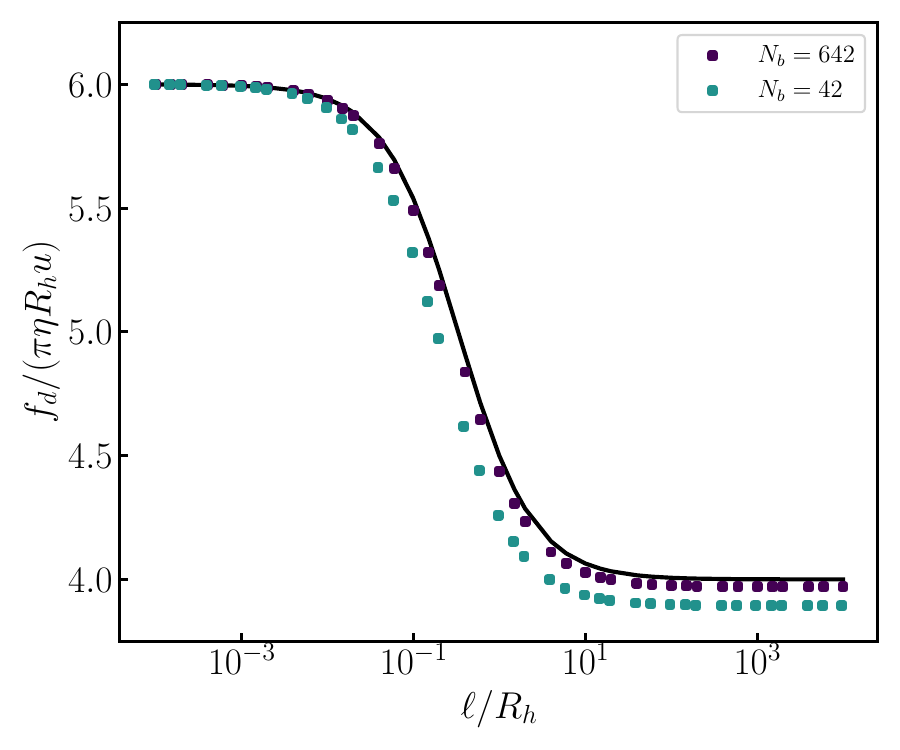}
\caption{
  Drag force $f_d$ on a sphere with different slip lengths, $\ell$, ranging from no-slip to full-slip.
  The curve represents the analytical result and the symbols the numerical results for two discretizations with $N_b=42$ and 642 blobs respectively.
  The numerical results are normalized with the hydrodynamic radius, $R_h$, computed in the no-slip case ($\ell / R \approx 10^{-4}$).
}
\label{fig:Rh_vs_slip}
\end{figure}

As second test to validate our partial slip BCs,
we solve a mobility problem to compute the drag force $f_d$ on a sphere with homogeneous slip but different slip length values $\ell$.
We show in Fig.\ \ref{fig:Rh_vs_slip} the numerical results for two discretizations with $N_b=42$ and $642$ respectively and the
analytical drag force $f_d = 6 \pi \eta R \left( \frac{1 + 2\ell/R}{1 + 3\ell/R} \right) u$ \cite{Happel1983,PADMAVATHI1993}.
We normalize our numerical results with the hydrodynamic radius, $R_h$, computed in the no-slip case.
For the coarser resolution with 42 blobs, the numerical error is below $3\%$ for all slip lengths while
for the fine resolution with 642 blobs is below $1\%$.
This showcases that low resolutions can be used to get accurate results up to a few percent error.

\subsection{Janus particles}
\label{sec:janus}

\begin{figure}
\centering
\includegraphics[width=0.99 \textwidth]{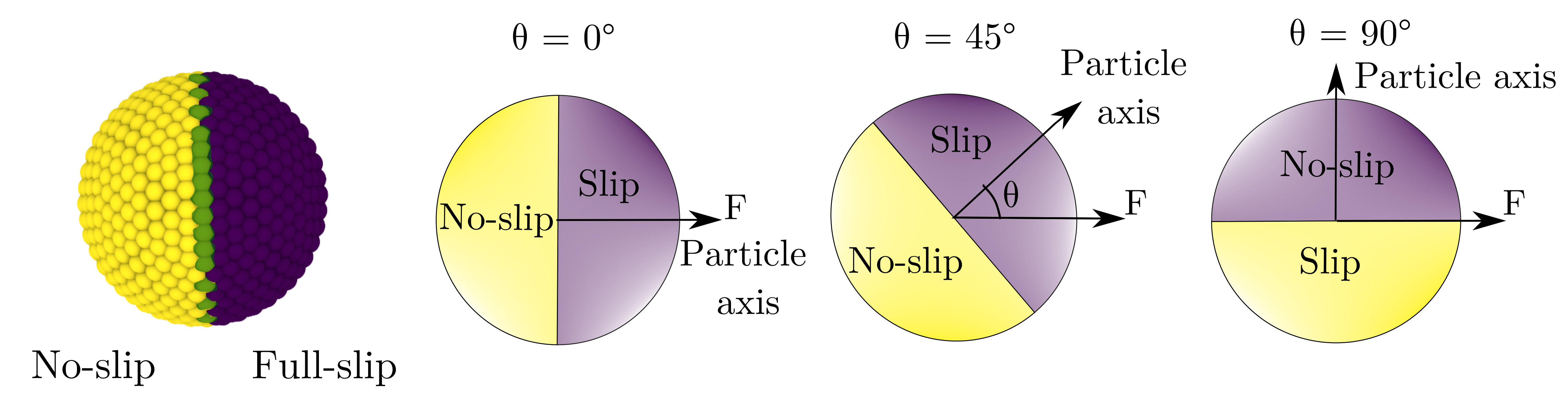}
\caption{Janus particles have one half (colored yellow) representing a no-slip surface, while the other half (colored purple) represents a full-slip surface.
  The interphase line region is an average slip length.
  The scheme illustrates examples of Janus particles aligned along the particle
  axis at three angles: 0, 45, and 90 degrees from the horizontal x-axis.
  In all cases, the applied force acts along the horizontal x-axis.  
}
\label{fig:janus_model}
\end{figure}

In the previous validations we tested spherical particles across a full range of slip conditions, from no-slip to full-slip, but always with an homogeneous slip.
However, the proposed scheme allows for definition of heterogeneous slip conditions over the surface of the particle, such can occur in decorated patchy nanoparticles or Janus particles.
Here, we measure the drag force and torque on a Janus particle following the work of Sun et al.\ \cite{Sun2013}.
On the Janus particle we apply non-homogeneous BCs: one half is set to full-slip ($\ell/R_g = 10^{4}$) while
the other half is set to no-slip ($\ell/R_g = 10^{-4}$) and an intermediate region is assigned a slip length that is the average of these two values, see Fig.\ \ref{fig:janus_model}.
Instead of solving the mobility problem, i.e.\ applying forces and torques to compute the body velocities, we solve a resistance problem.
We set the particle velocities to $\bu=(0,0,1)$ and $\bomega=\bzero$ and compute the resulting force-torque $\bF$ for different orientations
respect to the direction of motion as shown in Fig.\ \ref{fig:janus_model} \cite{Sun2013}.

\begin{figure}
\centering
\includegraphics[width=0.95 \textwidth]{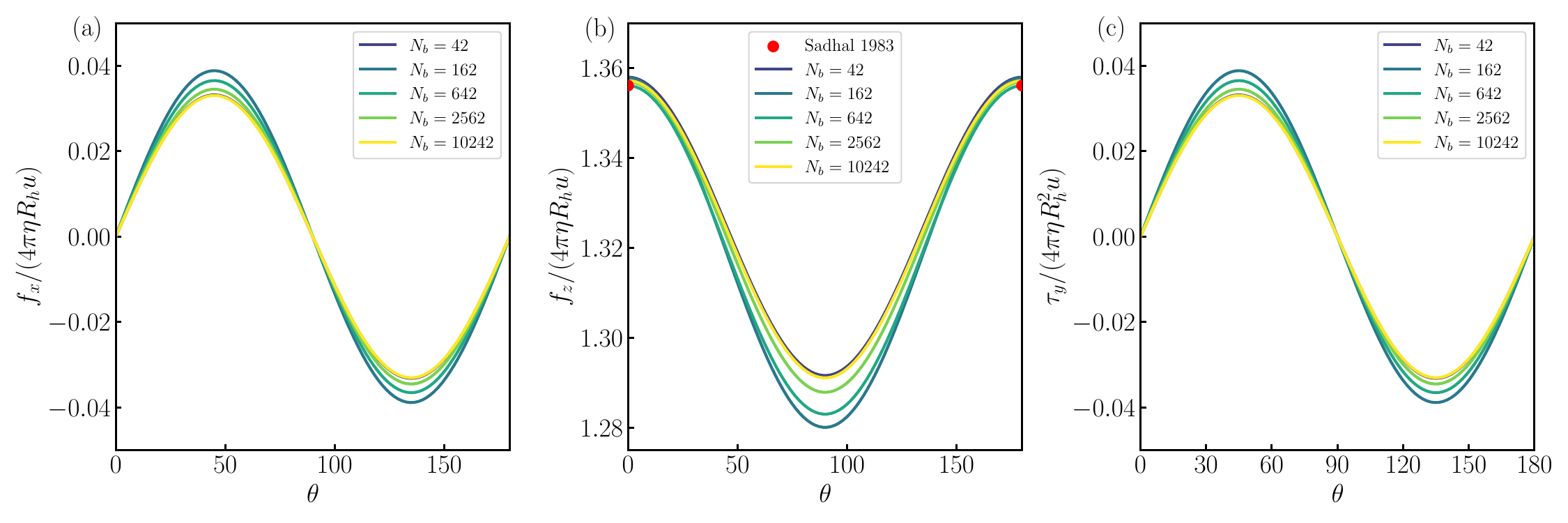}
\caption{Force ($f_z, f_x$) and torque ($\tau_y$) components
  for spherical Janus particle as functions of the angle $\theta$.
  Results for discretization with different numbers of blobs, $N_b$.
  The analytical result of Sahdal for $f_z$ at $\theta = 0, \pi$ is shown as a red point in panel (b) \cite{Sadhal1983}.
}
\label{fig:janus}
\end{figure}

Figure \ref{fig:janus} presents the drag force and torque for a Janus particle discretized with different number of blobs.
The results for $f_z$ converge to the analytical result of Sadhal and Johnson for $\theta=0 \text{ and } \pi$, see \ref{fig:janus}b \cite{Sadhal1983}.
For other angles or force-torque components our results show self-convergence.
We should mention that our results also agree to within $1\%$ to the results' of Sun et al.\ \cite{Sun2013} (not shown) which have an error of $1\%$ respect the
analytical result of Sadhal and Johnson.

\subsection{Random slip distribution}
\label{sec:random}

\begin{figure}
\centering
\includegraphics[width=0.9 \textwidth]{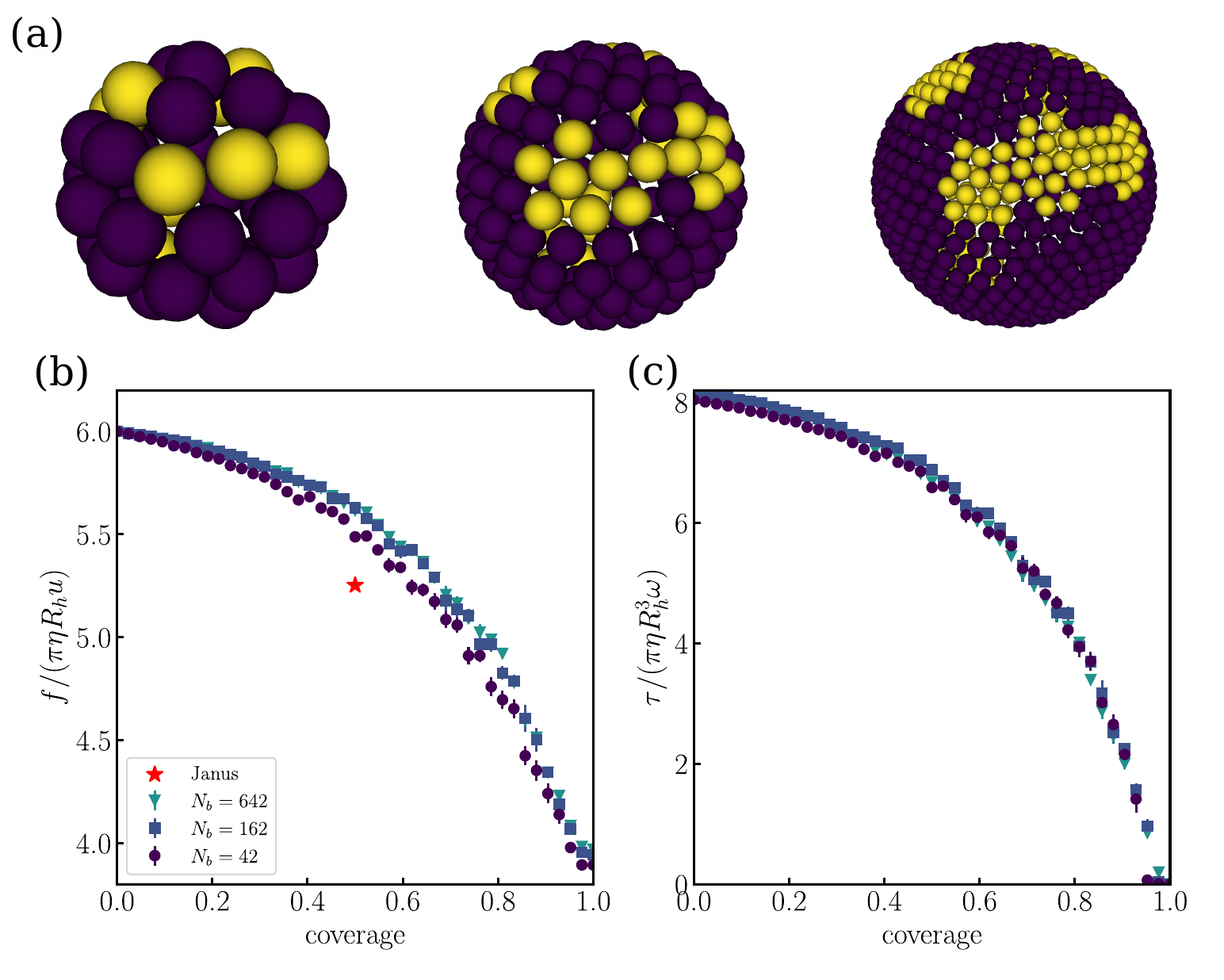}
\caption{
  {\bf (a)} Example of a sphere discretized with $N_b=42$ blobs and a random slip distribution and refined discretizations
  with 162 and 642 blobs and the same slip distribution.
  {\bf (b,c)} Translational and rotational drag versus slip coverage from full no-slip to full-slip computed for three resolutions.
  For each coverage 10 random slip distributions are used and the standard deviation in the drag is shown as error bars.
  The red star in {\bf(b)} represents the drag of a Janus sphere discretized with $N_b=10242$ blobs as in Sec.\ \ref{sec:janus}.
}
\label{fig:random_slip}
\end{figure}

The current method allows for more complex slip distributions.
To showcase the flexibility of the method, we measure the translational and rotational drag on spheres with random slip distributions.
We take a spherical particle discretized with $N_b=42$ blobs and set the slip on the blobs to either full-slip ($\ell/R=10^3$) or no-slip
($\ell/R=10^{-3}$), varying the full-slip coverage from 0\% to 100\%.
For each slip coverage we use 10 random slip distributions and measure the average drag and the standard deviation, see Fig.\ \ref{fig:random_slip}bc.
Interestingly, the results show a nonlinear variation between the full no-slip and full-slip distributions, while the drag standard deviations,
shown as error bars in Fig.\ \ref{fig:random_slip}bc, are small.
The translational drag standard deviation is below 5\% for all slip coverages.
For 50\% full-slip coverage spheres with random slip distributions have a drag about 10\% larger than a Janus sphere,
whose drag is shown as a red star in Fig.\ \ref{fig:random_slip}b.

Next, we study the convergence of the numerical results with increased resolution.
For each slip arrangement, we increase the resolution to 162 and 642 blobs while maintaining the same slip distribution, see Fig.\ \ref{fig:random_slip}a.
The drag computed with the refined resolutions agrees well with the coarse resolution.
Thus, coarse resolutions can be used as long as the blob radius is similar to, or smaller, than the characteristic length of the spatial slip distribution.

\subsection{Spherocylinders}
\label{sec:rods}

\begin{figure}
\centering
\includegraphics[width=0.9 \textwidth]{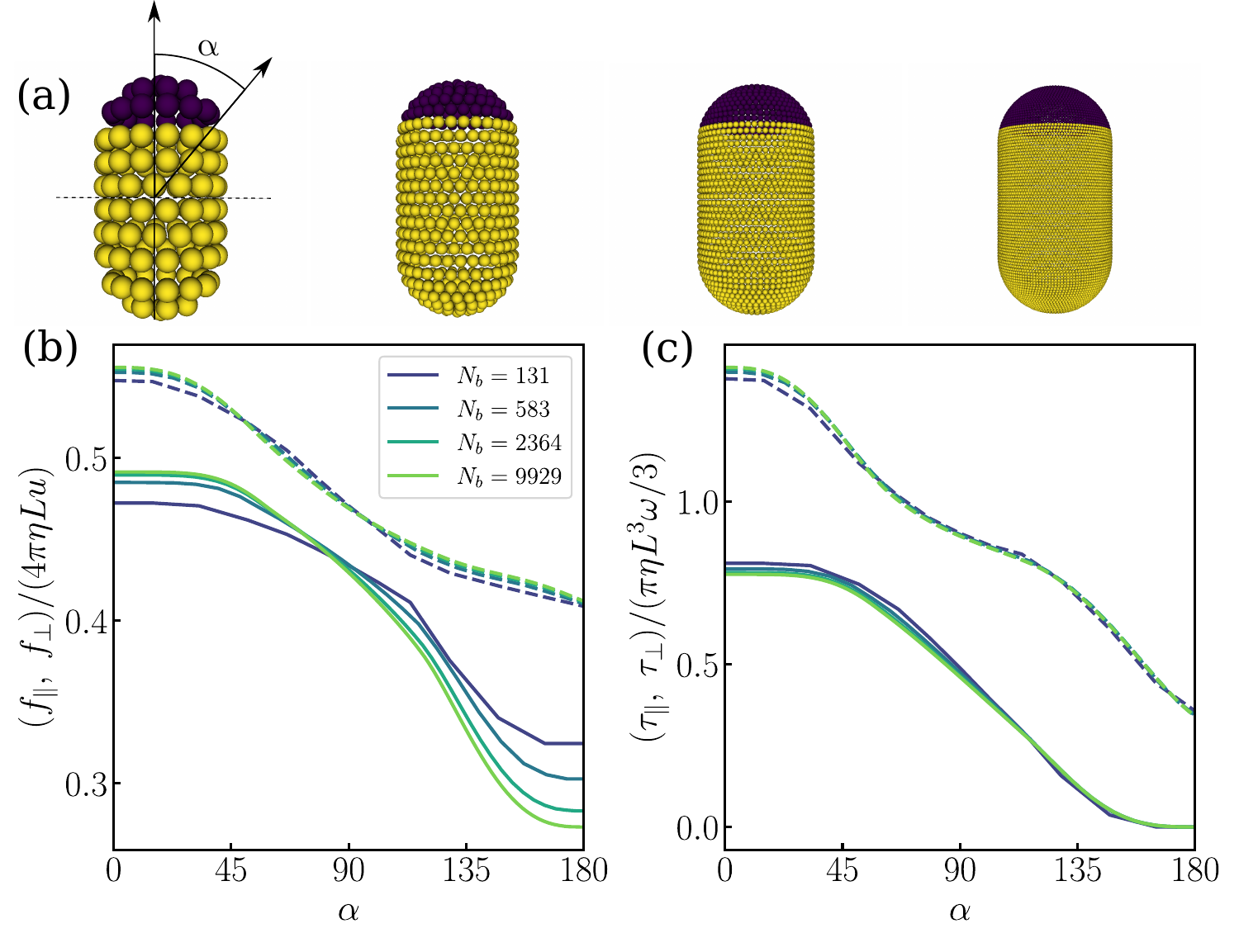}
\caption{
  {\bf (a)} Spherocylinders of aspect radio $4$ discretized (from left to right) with $N_b=132,\, 583,\, 2364,\text{ and } 9929$ blobs.
  The polar angle $\alpha$ sets the interface between the domain with full-slip (dark blobs) or no-slip (light blobs).
  {\bf (b)} Spherocylinder' translational drag parallel to its main axis (continuous lines) and perpendicular (dashed lines)
  versus the $\alpha$.
  {\bf (c)} Rotational drag, with the same notation as in (b).
}
\label{fig:mobility_rod}
\end{figure}

In this section, we study the capability of our numerical method to model spherocylindrical particles.
We use spherocylinders formed by a cylindrical body of length $L$ and radius $R=L/2$, with two spherical caps of radius $R$ at the ends.
First, we consider Janus particles with an axisymmetric slip length distribution.
The surface of a Janus particle is divided into two domains: one with no-slip ($\ell/L=10^{-3}$) and the other with full-slip ($\ell/L=10^{3}$).
The position of the interface between the two domains is set by the polar angle $\alpha$, see Fig.\ \ref{fig:mobility_rod}a.
We measure the translational and rotational drag as $\alpha$ varies smoothly from $\alpha=0$ to $180^{\circ}$,
i.e., from a particle with homogeneous no-slip boundary conditions to a particle with homogeneous full-slip boundary conditions.
To calculate the translational drag, we measure the force on a particle moving with unit velocity along, or perpendicular, to its main axis.
Similarly, we measure the rotational drag when it rotates with unit angular velocity around, or perpendicular, to its main axis.

We present the drag results obtained with resolutions between $N_b=132$ and $N_b=9929$ blobs in Fig.\ \ref{fig:mobility_rod}bc.
All the drag components decay with $\alpha$, i.e., with increasing full-slip coverage.
The largest change appears in the rotational drag around the main axis.
In that case, the drag decays to zero for a homogeneous full-slip rod because all the fluid motion is tangential to the surface.
In the other cases, the flow has non-tangential components, and thus the drag never decays to zero, even for the homogeneous full-slip case.
The results show that a coarse resolution of $N_b=132$ blobs is sufficient to obtain results accurate within 10\% compared to
the refined results for all cases except for the homogeneous full-slip case, where the errors are around 15\% for the translational drag along the main axis.
Interestingly, the rotational drags have small errors for all slip distributions.

\begin{figure}
\centering
\includegraphics[width=0.9 \textwidth]{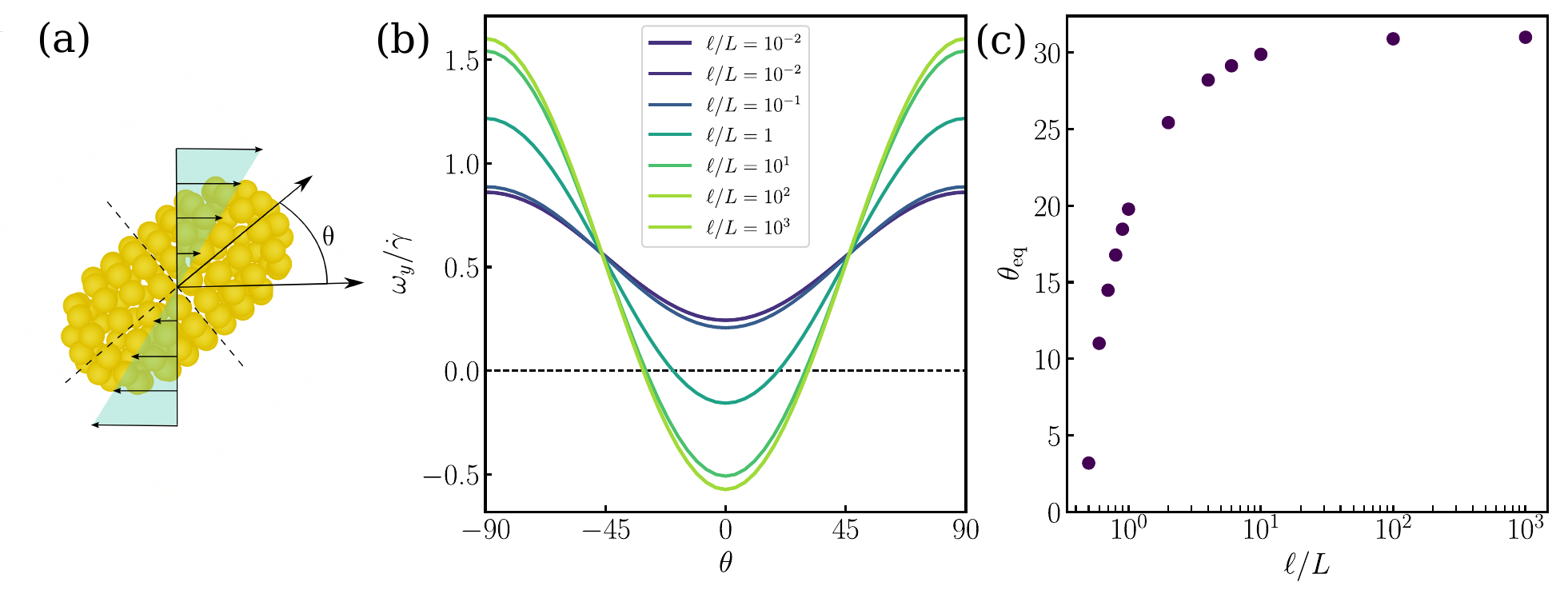}
\caption{
  {\bf (a)} Sketch of a spherocylinder in a shear flow and definition of $\theta$, the orientation angle with the flow.
  {\bf (b)} Angular velocity of a spherocylinder in a shear flow respect to the angle $\theta$ formed
  by the cylinder main axis and the direction of the flow.
  For small slip lengths the angular velocity is always positive while for large slip lengths the angular velocities
  become negative for particles nearly aligned with the shear flow.
  {\bf (c)} Equilibrium angle with the shear flow respect the slip length.
}
\label{fig:wy}
\end{figure}

As a second test, we study how the angular velocity of a spherocylinder with the same geometry ($R=L/2$)
and a homogeneous slip distribution, immersed in a shear flow $\bv_0 = \dot{\gamma} z \be_x$, varies with the slip length.
Since we showed in the previous test that the rotational drag was recovered with good accuracy using a coarse discretization,
in this test we use a particle discretized with $N_b=132$ blobs, see Fig.\ \ref{fig:wy}a.
We show in Fig.\ \ref{fig:wy}b the angular velocity of the particle with respect to its orientation with the flow for several slip lengths.
For small slip lengths, the angular velocity is always positive, i.e., the particle rotates with the flow as expected.
Interestingly, for slip lengths larger than $\ell/L \approx 1/2$, the angular velocity becomes negative for some orientations,
i.e., the particle counter-rotates with respect to the flow.
The sign flip in the angular velocity indicates that particles with sufficiently large slip lengths can reach
an equilibrium orientation with respect to the shear flow.
We show in Fig.\ \ref{fig:wy}c the equilibrium angle for several slip lengths.
The equilibrium angle increases monotonically with $\ell$ and reaches a plateau for $\ell/L \approx 100$.
This curious phenomenon, where a particle aligns in a shear flow, has been observed in graphene nanoplatelets, which exhibit
very large slip lengths \cite{Kamal2020, Kamal2021}.
We show here that the same effect appears for rod-like particles with sufficiently large slip lengths.

\subsection{Suspension viscosity}
\label{sec:viscosity}

In this section we measure the high frequency viscosity in a suspension of spherical particles with homogeneous slip length and
we explore how the suspension viscosity changes with volume fraction and slip length.
We study suspensions with 250 particles in a triply periodic domain of size $L$ which we vary to set the desired volume fraction $\phi$.
In this section we use particles discretized with $N_b=162$ blobs.
To measure the high frequency shear viscosity we follow the following protocol.
First, we generate 100 random configurations using a Monte Carlo code.
Then, for each configuration we apply a sinusoidal background flow
\eqn{
  \bv_0(x,y,z) = v_0  \sin\pare{\fr{2\pi z}{L}} \be_x,
}
and we measure the magnitude of the flow velocity profile $v_x$. Then, we extract the viscosity from the flow magnitude as
\eqn{
  \label{eq:viscosity}
  \eta_{\text{susp}} = \fr{v_x}{v_0}.
}

However, to obtain accurate results some care it is necessary in applying the background flow and in measuring the resulting flow.
As the Stokes equations are linear the background flow can be included in the right hand side of the slip equation, that is
\eqn{
  \label{eq:slip_eq_with_v0}
  \bM \blambda - \pare{\fr{1}{2}\bI + \bD}\bK\bU - \pare{\fr{1}{2}\bI + \bD} \bu_s = -\bv_{0}.
}
However, we obtained results with large numerical errors using this straightforward approach.

In order to improve the accuracy it is instructive to consider what happens when one introduces a constant background flow, e.g.\ $\bv_0 = \be_x$.
In the continuum limit the action of the double layer operator on rigid velocities is equivalent to
half the identity operator, i.e.\ $\pare{\fr{1}{2}\bmI + \bmD}\bmK\bU = \bmK\bU$ \cite{Pozrikidis1992}.
Thus, for constant background flow the solution of Eq.\ \eqref{eq:slip_eq_with_v0} is simply $\bu=\bv_0$, i.e.\ the particles are advected by the flow as expected.
However, in the discrete setting the above relation only holds approximately,
in fact $\pare{\fr{1}{2}\bI + \bD}\bK\bU = \bK\bU + \mc{O}\pare{a}$ where the error term is proportional to the blob radius $a$.
Thus, the particles are advected with a different velocity than the background flow even in this simple example.
A similar error occurs with other background flows and that error is transmitted to the viscosity measurement.

\begin{figure}
  \centering
  \includegraphics[width=0.5 \textwidth]{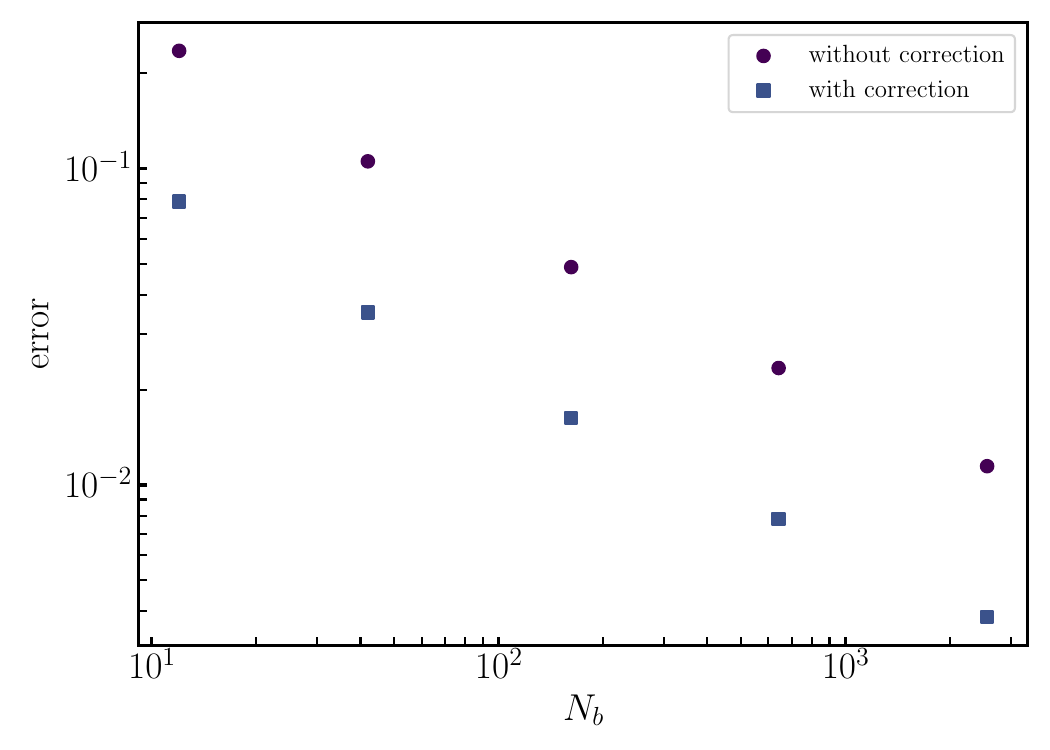}
  \caption{Error in the angular velocity of a single spherical particle immersed in
    a shear flow versus the number of blobs $N_b$ used to discretize the sphere
    with and without the background flow correction Eq.\ \eqref{eq:v_RHS}.
    With the correction the error is about three times smaller.
  }
\label{fig:error}
\end{figure}

Our strategy to reduce the numerical error is to add a correction term to the background flow that vanishes in the continuum limit
(number of blobs $N_b \rightarrow \infty$) but that guarantees that particles are advected with the correct velocity for constant background flows
for all resolutions.
These two conditions are satisfied by adding and subtracting a rigid body contribution to the background flow,
so in the right hand side of Eq.\ \eqref{eq:slip_eq_with_v0} we use the \emph{effective} flow
\eqn{
  \label{eq:v_RHS}
  \bv_{\text{RHS}} = \bv_0 - \bK \pare{\fr{1}{N_b} \bK^T \bv_0} + \pare{\fr{1}{2}\bI + \bD}  \bK \pare{\fr{1}{N_b} \bK^T \bv_0}.
}
In the continuum limit, the last two terms in Eq.\ \eqref{eq:v_RHS} cancel out, thus the continuum problem is not modified.
In the discrete case, for constant background flow,
the first two terms cancel exactly and the last term balances the term $\pare{\fr{1}{2}\bI + \bD}\bK\bU$ in Eq.\ \eqref{eq:slip_eq_with_v0}
which guarantees that particles are advected with the background velocity to machine precision.
This correction also reduces the error in the angular velocity of a single spherical particle immersed in a shear flow
as shown in Fig.\ \ref{fig:error}.
Using this background flow it is possible to obtain accurate viscosity results as we will show in a moment.

The second detail to extract the viscosity with higher accuracy is to measure the flow velocity correctly.
Since we solve the Stokes equations with an implicit solvent method, we do not have direct access to the flow (although it can be computed).
Therefore, we measure the average particle velocities profile, $\avg{u_x}(z)$, and use the Faxén law to extract the ambient velocity magnitude from
\eqn{
  \avg{u_x} = \pare{\bI + \fr{R_h^2}{6}\bna^2}v_x \cos\pare{\fr{2\pi z}{L}}.
}
The flow magnitude, $v_x$, extracted from this equation is used in Eq.\ \eqref{eq:viscosity} to compute the suspension velocity.

\begin{figure}
  \centering
  \includegraphics[width=0.95 \textwidth]{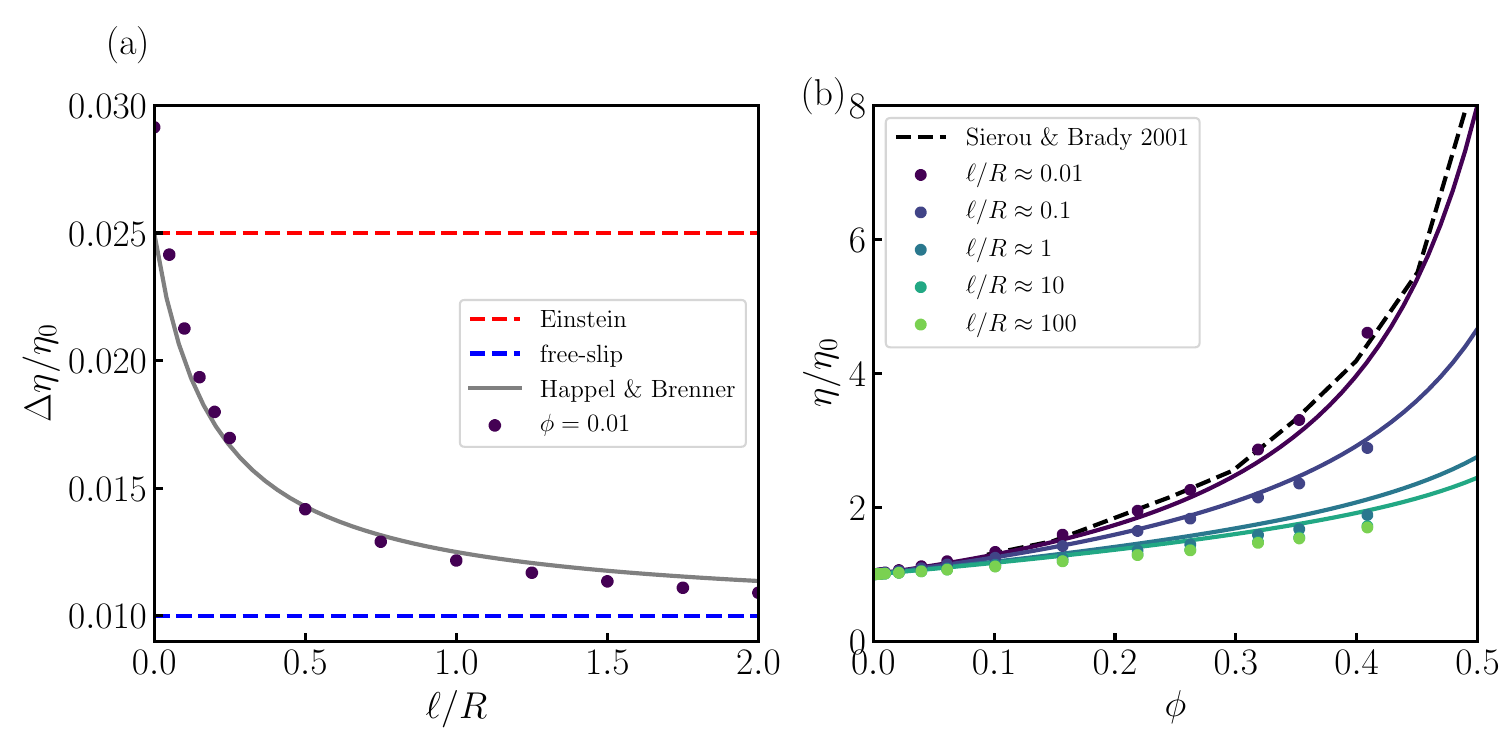}  
  \caption{High frequency viscosity for suspension of spherical particles.
    (a) Relative viscosity increase at low volume fraction ($\phi = 0.01$) versus slip length.
    (b) Relative viscosity versus volume fraction for several slip lengths.
    Symbols represent numerical results, continuous lines the theoretical prediction of Ref.\ \cite{Housiadas2020}
    and dashed lines no-slip Stokesian results \cite{Sierou2001}.
  }
\label{fig:viscosity}
\end{figure}

In the left panel of Fig.\ \ref{fig:viscosity} we show the viscosity at low volume fraction, $\phi=0.01$, versus the slip length.
We compare the numerical results with analytical result from Happel and Brenner \cite{Happel1983}
\eqn{
  \label{eq:eta_Brenner}
  \fr{\eta_{\text{eff}}}{\eta} = 1 + \fr{5}{2}  \cdot \fr{1 + 2 \ell/R}{1 + 5 \ell/R} \phi,
}
which go from the no-slip limit (Einstein's limit) to the full-slip limit \cite{Taylor1932, Happel1983}. 
We observe a good agreement between numerical and theoretical results for all the slip lengths considered.

Finally, in the right panel of Fig.\ \ref{fig:viscosity} we show the suspension viscosity versus volume fraction for several slip lengths.
The no-slip results agree well with the Stokesian dynamics results of Sierou and Brady \cite{Sierou2001}, while
the viscosity grows slower for larger slip lengths as expected.
The viscosity decreases with the slip length until $\ell / R \approx 10$.
Further increases in the slip length do not reduce the viscosity any lower.
We also compare our numerical results with an approximated theory derived from a Stokes-Darcy model with partial slip \cite{Housiadas2020}.
For small slip lengths (e.g., $\ell \le 0.1$), the numerical results show good agreement with the analytical predictions.
However, for larger slip lengths, the theory proposed in Ref.\ \cite{Housiadas2020} consistently overpredicts the effective viscosity.
This overprediction occurs across all volume fractions, as the theory's prediction for low concentrations of free-slip particles,
$\eta_{\text{eff}}/\eta = 1 + 1.6\phi$, exceeds the exact prediction, to linear order, given by Eq.~\eqref{eq:eta_Brenner}.
The theory from Ref.\ \cite{Housiadas2020} derives the viscosity modeling the flow near a particle
with a Stokes-Darcy model and perhaps the additional dissipation introduced by the porosity of the Darcy model overpredicts the viscosity.
Nevertheless, the theory successfully captures the overall trend observed in the numerical results.

\section{Conclusions}
\label{sec:conclusions}

Deviations from the no-slip BCs can be modeled using partial slip BCs \cite{Bocquet2007, Neto2005}. 
This paper introduces a numerical method for simulating particle suspensions with partial slip BCs.
Our approach is based on a regularized boundary integral method. It is easy to implement, accurate within a few percent, and capable of simulating large-scale suspensions. The scheme employs a regularized double-layer formulation to implement the partial slip BCs \cite{Kamal2020, Kamal2021}. 
Only the particle surfaces are discretized within this framework, significantly reducing the algorithm’s memory requirements.
To solve the Stokes equations, we use a preconditioned GMRES solver.
With a simple block-diagonal preconditioner, the number of iterations required to solve the Stokes equations depends weakly on the number of particles
or the resolution used (see Figs.~\ref{fig:convergence_one_shell}--\ref{fig:convergence_many_shells}).

We have validated the accuracy of our scheme across a range of slip lengths, from full-slip to no-slip BCs. In all cases, our results remained accurate within a few percent, even with resolutions as low as 42 blobs per particle. Additionally, we have reproduced the drag forces and torques on Janus particles.
We have also measured the high-frequency viscosity in suspensions of particles with homogeneous slip distributions.
At low volume fractions, our viscosity results are in good agreement with analytical predictions \cite{Happel1983}.
Our results in the limit of small slip lengths ($\ell / R \approx 0.01$)
are consistent with no-slip Stokesian Dynamics simulations over a broad range of volume fractions \cite{Sierou2001}.
Then, as the slip length increases, the viscosity decreases as expected.
We found that the viscosity decreases with the slip length up to $\ell / R = 10$, beyond which it becomes independent of the slip length.

The use of an implicit solvent method reduces the memory requirements considerably as only the particles surfaces are discretized. 
For example, in the viscosity measurements of Sec.\ \ref{sec:viscosity} we employed 250 particles each of then discretized 
with 162 blobs, thus, $\num[group-separator={,}]{40500}$ blobs in total. 
This number of degrees of freedom is independent of the volume fraction, $\phi$, as the solvent itself is not explicitly discretized. 
With those degrees of freedom one mobility problem was solved within 6 to 20 iterations to a tolerance $10^{-6}$ in about 
one minute per time step using only 8 cores per simulations. 
Thus, it is possible to run large simulations in a single workstation. 
The same system solved with an explicit solvent method would have required 
about $(L / (2a))^3$ particles to discretize the system, where $L$ is the system size and $a$ the 
blob radius or equivalently the discretization length scale.
Using the same resolution as in our simulations ($R / a \approx 8$) that would be between $1.5 \cdot 10^5$ and $6\cdot 10^6$ particles 
for volume fractions between $\phi=0.4$ and $\phi=0.01$, like the ones used in Fig.\ \ref{fig:viscosity}. 
Thus, low volume fraction simulations become particularly expensive in explicit solvent methods.
On the other hand, inertial effects or non-Newtonian solvent properties require a full description of
the solvent domain \cite{Krishnan2017, VazquezQuesada2019}.

The scheme presented in this work is flexible enough to model inhomogeneous slip length distributions,
incorporating for instance, regular and irregular patterns over the surfaces,
that have potential applications in nanotechnology and biomedicine \cite{SU2019,Safaie2020,Palacios-Alonso2023}. 
The extension of this work to Brownian suspensions would be an important contribution.
However, how to generate Brownian velocities when the double layer operator, which is non-symmetric and thus not positive definite,
appears in Eq.\ \eqref{eq:slip_integral} remains an open problem.
For no-slip boundary conditions it is possible to generate Brownian displacements by
including in the right-hand-side of the mobility problem (Eq.\ \eqref{eq:linear_system}) 
a random flow compatible with the Stokes equations \cite{Sprinkle2017}.
The physical interpretation is that the particles diffuse by the motion generated by the random flow.
Using the fact that the mobility matrix $\bM$ is positive definite, it is possible to prove that this approach
obeys the fluctuation-dissipation balance \cite{Sprinkle2017}.
This suggest that the same idea should work for partial slip boundary conditions,
however, the presence of double layer operator in the mobility problem breaks the fluctuation-dissipation balance.
Bedeaux, Albano and Mazur studied the formulation of Brownian dynamics with non-zero slip lengths \cite{Bedeaux1977}.
They showed that in the context of fluctuating hydrodynamics it is necessary to include an
additional random frictional force at the fluid-particle interface to balance the dissipation generated by the slip
and fulfill the fluctuation-dissipation balance \cite{Bedeaux1977}.
Further exploration of this approach will be left for future work. 
Our implementation of the method is publicly available as free software at \url{https://github.com/stochasticHydroTools/RigidMultiblobsWall}.

\section*{Acknowledgments}
We acknowledge funding by the Basque Government through the BERC 2022-2025 program and by the Ministry of Science, Innovation and Universities:
BCAM Severo Ochoa accreditation CEX2021-001142-S/MICIN/AEI/10.13039/501100011033.
The Spanish State Research Agency through the project PID2020-117080RB-C55 MICIU/AEI /10.13039/501100011033
funded by (AEI/FEDER, UE) with acronym COMPU-NANO-HYDRO.\\\\


Declaration of generative AI and AI-assisted technologies in the writing process:\\
During the preparation of this work the author(s) used Chat GPT and Grammarly in order to check the grammar and readability of the text. After using this tool/service, the author(s) reviewed and edited the content as needed and take(s) full responsibility for the content of the publication.

\bibliographystyle{elsarticle-num}
\bibliography{biblio.bib}

\end{document}